\documentstyle[a4,12pt,graphicx]{article}

\newcommand{\ee}{\end{equation}}
\newcommand{\be}{\begin{equation}}

\newcommand{\R}{{\rm I \hspace{-0.9mm} R}}
\newcommand{\CE}{{\cal E}}
\newcommand{\CL}{{\cal L}}
\newcommand{\CH}{{\cal H}}

\newcommand{\sPhi}{\bar{\Phi}}
\newcommand{\sA}{\bar{A}}

\newcommand{\tr}{\mbox{tr}}
\newcommand{\Ncs}{{\cal N}_{CS}}

\renewcommand{\d}{{\rm d}}

\begin{document}

\title{Sphaleron of a 4 dimensional $SO(4)$ Higgs model}
\author{{\large B. Kleihaus}$^{\dagger}$
{\large D.H. Tchrakian}$^{\dagger \star}$
and {\large F. Zimmerschied}$^{\dagger}$ \\ \\
$^{\ddagger}${\small Department of Mathematical Physics, National
University of Ireland Maynooth,} \\
{\small Maynooth, Ireland}\\ \\
$^{\star}${\small School of Theoretical Physics -- DIAS, 10 Burlington Road,
Dublin 4, Ireland }}

\maketitle
\vspace{5cm}
\thispagestyle{empty}
\begin{abstract}
We construct the finite energy path between topologically distinct vacua
of a 4 dimensional $SO(4)$ Higgs model which is known to support an
instanton, and show that there is a sphaleron with Chern--Simons number
$\Ncs={1\over 2}$ at the top of the energy barrier. This is carried out
using the original geometric loop construction of Manton.
\end{abstract}
\medskip
\medskip
\newpage

\section{Introduction}

The basic $SO(4)$ Higgs model~\cite{OT1} in 4 Minkowskian dimensions supports
instanton solutions in Euclidean spacetime, that are evaluated 
numerically~\cite{AOT}. This
model is arrived at by dimensional descent from the 8 dimensional
Yang--Mills action~\cite{T} stabilised by the 4-th Chern class. It is
also expected that in the static limit this system supports sphaleron
solutions, as described in Ref.~\cite{OT2} where such solutions were
however not evaluated numerically.

Subsequently, in a different context,
monopole solutions to a generalised $SO(3)$ Higgs model (generalising
the Georgi--Glashow model in 3 dimensions) were numerically~\cite{KOT}
evaluated, which happen to describe the sphaleron solution of the 4
dimensional $SO(4)$ Higgs model concerning us in the present work. We
will explain this connection here and will construct the finite energy
path between two topologically distinct vacua describing the geometric 
non contractible loop (NCL)~\cite{Manton}, with the sphaleron of 
Chern--Simons number $\Ncs={1\over 2}$ at its top. These are the results
reported in the present note. Before proceeding to these technical tasks,
we note some properties
of this model and its extensions, which may be of some physical relevance.

There are two distinct physical justifications for studying this model and
some extensions. The first is that some extended versions of it, which
certainly~\cite{AOT} support instantons, are capable of describing a
Coulomb gas of instantons. This allows the attempt at extending the work
of Polyakov~\cite{P} exploiting a Coulomb gas of instantons in {\it three}
dimensions, to {\it four} dimensions. (Work in this direction is under
active consideration.) The second physical relevance of this model is in
the fact that it supports both instantons and sphalerons, in contrast
for example to the standard electroweak model which supports only the
latter, and whose instantons have shrinking size unless if constrained
instantons~\cite{A} are employed. It is known that theories which
support {\it both} finite size instantons {\it and} sphalerons on the
one hand, and those which support {\it only} sphalerons on the other,
have quite distinct properties concerning the contribution of periodic
instantons~\cite{KRT} relative to that of the sphaleron and the instanton.
This situation has been clearly demonstrated in Ref.~\cite{KT}, in the
case of various (1+1) dimensional $O(3)$ sigma models with the above
described relative properties.

In Section 2, we present the model and explain how the generalised
monopole~\cite{KOT} gives the sphaleron solution. This is done using a
``geometrical'' 
NCL Ansatz~\cite{Manton}. In Section 3, we construct the finite energy
barrier as a NCL with the sphaleron at the top. Section 4 is devoted to a
discussion of our results in the context of some extensions of the model
used here, which help to highlight the physical relevance of these results
and point out the theoretical obstacles that must be tackled.

\section{The model and its properties}

We consider a model in $d=4$ spacetime dimensions,
consisting of a $SO(4)$ gauge field and a Higgs quartet
field $\phi^a$. Using Euclidean gamma matrices to represent both the
gauge and the Higgs field, the gauge field takes its values
in the $so(4)$ algebra with generators 
$\gamma_{\mu\nu}=-\frac14[\gamma_{\mu},\gamma_{\nu}]$
and the Higgs field is written in antihermitean isovector matrix 
representation, $\Phi=\phi^a\gamma_5\gamma_a$.

As we are interested in the instanton and (particularly) sphaleron physics of 
the model, we give its Lagrangian \cite{OT1} in Euclidean signature,
\begin{equation}
\CL = \tr\left(S_{\mu\nu\rho\sigma}^2+
4\lambda_4S_{\mu\nu\rho}^2+18\lambda_3S_{\mu\nu}^2+54\lambda_2S_{\mu}^2
+54\lambda_1S^4\right)
\label{m1}
\end{equation}
with
\begin{eqnarray}
S_{\mu\nu\rho\sigma} & = & \{F_{\mu[\nu},F_{\rho\sigma]}\} \nonumber \\
S_{\mu\nu\rho} & = & \{F_{[\mu\nu},D_{\rho]}\Phi\} \nonumber \\
S_{\mu\nu} & = & i\left(\{S,F_{\mu\nu}\}+[D_{\mu}\Phi,D_{\nu}\Phi]\right)
\nonumber \\
S_{\mu} & = & i\{S,D_{\mu}\Phi\} \nonumber \\
S & = & -(\Phi^2+\eta^2)
\label{m1a}
\end{eqnarray}
The curvature is given by 
$F_{\mu\nu}=\partial_{[\mu}A_{\nu]}+[A_{\mu},A_{\nu}]$, and the
covariant derivative is defined by
$D_{\mu}\Phi = \partial_{\mu}\Phi + [A_{\mu},\Phi]$.
The coupling constants are assumed to be positive, $\lambda_a>0$. The
instanton solutions to this system were evaluated numerically in
Ref.~\cite{AOT}. Their salient properties are that they are exponentially
localised and, the connection is {\it not} asymptotically pure gauge,
both in contrast to the BPST~\cite{BPST} instanton.
The first of these properties results in the finite size of the instanton,
fixed by the absolute scale $\eta$.
The second property results because the asymptotic gauge field decays as 
one half times a pure gauge, rather 
like a 't~Hooft--Polyakov monopole~\cite{tHP}. A very important
consequence is that the curvature field of this instanton features an
inverse--square decay, leading to the possibility of constructing a
Coulomb gas of instantons~\cite{AOT}.

The existence of finite size instantons ensures that the basic model (\ref{m1})
has topologically distinct vacua separated by energy barriers of finite height.
To construct a path connecting two vacua ``over the barrier'' in the space of
static configurations with finite energy
$\CE=\int \CH\,d^3x$ \cite{OT2} where $\CH$ is the static Hamiltonian
\begin{equation}
\CH= \tr\left(
4\lambda_4S_{ijk}^2+18\lambda_3S_{ij}^2+54\lambda_2S_{i}^2
+54\lambda_1S^4\right),
\label{m6}
\end{equation}
we follow a standard geometrical technique \cite{Manton} exploiting the
topological properties of the model, resulting from the requirement of
finite energy.

The Higgs field of any static finite energy configuration has to satisfy 
$|\Phi|^2=\eta^2$ at spatial infinity $(\R^3)^{\infty}=S^2_{space}$, 
hence any finite energy configuration defines a mapping 
$\Phi^{\infty}:S^2_{space}\rightarrow S^3_{Higgs}$. This
property allows to consider a one parameter set of static field 
configurations $\sPhi$ parameterised by a {\em loop parameter} 
$\tau\in[0,\pi]:=I_{Loop}$ such that these fields at 
spatial infinity define a topologically nontrivial mapping
\begin{equation}
\sPhi(r\rightarrow\infty):=
\sPhi^{\infty}:S^2_{space}\times I_{Loop}\sim S^3\rightarrow S^3_{Higgs}
\label{m7}
\end{equation}
which maps $S^2_{space}$ to $S^3_{Higgs}$ for any fixed value of 
$\tau\in(0,\pi)$. A simple, geometrically motivated \cite{Manton}
choice of such a mapping is given by
\begin{equation}
\sPhi^{\infty}(\tau,\theta,\phi)=\eta\gamma_5\vec{\gamma}\cdot
\vec{p}(\tau,\theta,\phi), \qquad 
\vec{p}(\tau,\theta,\phi):=\left(\begin{array}{c}
\sin\tau\sin\theta\cos\phi \\
\sin\tau\sin\theta\sin\phi \\
\sin^2\tau\cos\theta+\cos^2\tau \\
\sin\tau\cos\tau(\cos\theta-1)\end{array}\right).
\label{m8}
\end{equation}

Finite energy also requires the covariant derivative to vanish at spatial 
infinity. This fixes the gauge fields $\sA_i$, $\sA_4\equiv 0 $ (temporal 
gauge) along the loop at infinity to
\begin{equation}
\sA_i(r\rightarrow\infty)=:
\sA_i^{\infty}=-\frac{1}{4\eta^2}[\sPhi^{\infty},\partial_i\sPhi^{\infty}]
=-p^{\mu}p^{\nu}_i\gamma_{\mu\nu}
\label{m9}
\end{equation}
with $\vec{p}_i=\partial_i\vec{p}$.

The loop itself has to start ($\tau=0$) and end ($\tau=\pi$) in the vacuum
which we choose to be $\Phi^{(V)}=\eta\gamma_5\gamma_3$, $A_i^{(V)}=0$.
This allows the gauge field along the loop to be chosen proportional to the 
gauge field at infinity, introducing a radial profile function $f(r)$,
whereas the Higgs field has to be deformed to reach the vacua:
\begin{equation}
\sPhi=h(r)\sPhi^{\infty}+(1-h(r))\Psi, \qquad \sA_i=(1+f(r))\sA_i^{\infty}
\label{m10}
\end{equation}
with $\Psi=\eta\gamma_5[\gamma_3\cos^2\tau-\gamma_4\sin\tau\cos\tau]$.
The profile functions $h(r)$ and $f(r)$ are subject to the boundary conditions
\begin{equation}
h(0)=0,\quad f(0)=-1;\qquad
h(r\rightarrow\infty)=1,\quad f(r\rightarrow\infty)=0,
\label{m11}
\end{equation}
resulting from the requirements of regularity at the origin and finite energy.
The loop resulting from this construction is noncontractible as 
$\sPhi^{\infty}$
was chosen to be a topologically nontrivial mapping.

Inserting the ansatz into the static Hamiltonian (\ref{m6}) and multiplying
by the radial integration measure, we obtain the radial subsystem
\begin{eqnarray}
H_{(\tau)}[h,f]
& = & 96\pi\eta^6\sin^4\tau\Bigg[4\lambda_4
\left\{\frac{1}{\rho^2}\left(\left[(1-f^2)h'+2f'W\right]^2\sin^2\tau
+\left[2f'W\right]^2\cos^2\tau\right)\right\} \nonumber \\
& & \qquad \qquad\:\: {} + 
6\lambda_3\Bigg\{2\left[(1-h^2)f'\sin^2\tau+2h'W\right]^2+
2\left[(1-h^2)f'\right]^2\sin^2\tau\cos^2\tau \nonumber\\
& & \qquad \qquad\qquad\qquad 
{} + \frac{1}{\rho^2}\left[(1-h^2)(1-f^2)\sin^2\tau+2W^2\right]^2
\Bigg\} \nonumber\\
& & \qquad \qquad\: {} 
+36 \lambda_2 \left\{(1-h^2)^2\left[(\rho h')^2+2W^2\right]\sin^2\tau
\right\} \nonumber \\
& & \qquad \qquad\: {} 
+\:\: 9\lambda_1 \left\{\rho^2(h^2-1)^4\sin^4\tau\right\}
\label{m12}
\end{eqnarray}
with
\begin{equation}
W:=h-(1+f)(\cos^2\tau+h\sin^2\tau).
\label{m13}
\end{equation}

It should be emphasised that extrema of the radial subsystem (\ref{m12}) are 
not necessarily extrema of the static Hamiltonian (\ref{m6}) as the ansatz
(\ref{m10}) is in general not spherically symmetric. Nevertheless, besides
the (spherically symmetric) vacua for $\tau=0$ and $\tau=\pi$, 
the loop ansatz reduces to a spherically symmetric ansatz for 
$\tau=\frac{\pi}{2}$,
\begin{equation}
\sPhi\big|_{\tau=\frac{\pi}{2}}=\eta h(r)\gamma_5\gamma_i\hat{x}_i,\qquad\qquad
\sA_i\big|_{\tau=\frac{\pi}{2}}=\frac{1+f(r)}{r}\gamma_{ij}\hat{x}_j.
\label{m14}
\end{equation}
For this value $\tau=\frac{\pi}{2}$, the radial subsystem loop Hamiltonian
(\ref{m12}) reduces to the radial subsystem Hamiltonian of the generalised 
$SO(3)$ Higgs monopole system \cite{monopole} (the coupling constants have to 
be adjusted to $\lambda_a\mapsto \frac{1}{a}\lambda_a$, $a=1,2,3,4$)
for which solutions $(h^{(M)},f^{(M)})$ are known numerically. 

\begin{figure}
\begin{center}
\includegraphics[bb=3cm 10cm 18cm 21cm,angle=0,scale=0.8]{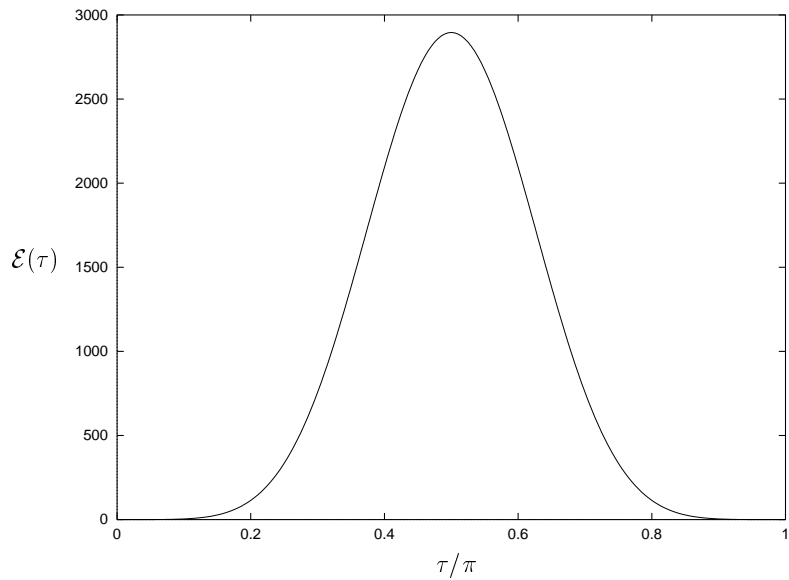}
\end{center}
\caption{Energy along the geometrical loop in terms of the loop parameter
$\tau$, $\lambda_a=\frac{1}{a}$, $\eta=1$}
\label{so4fig1}
\end{figure}

Due to spherical symmetry, inserting
these monopole profile functions into the ansatz (\ref{m14}) yields an extremum
of the static energy functional (\ref{m6}). Since by construction this 
extremum appears along a path connecting two vacua, it is expected to be
a saddle point. This becomes manifest if one inserts the monopole 
profile functions $(h^{(M)},f^{(M)})$ into the loop Hamiltonian (\ref{m12})
and considers the energy 
$\CE(\tau)=\int H_{(\tau)}[h^{(M)},f^{(M)}]dr$ along the loop which can be 
calculated numerically using the data known from the monopole analysis
\cite{monopole}, e.g.\ for coupling constants $\lambda_a=\frac{1}{a}$,
$a=1,2,3,4$.
Fig.\ \ref{so4fig1} shows that for this loop, $\CE(\tau)$ really reaches 
its maximum for $\tau=\frac{\pi}{2}$ which proves that the extremum we found 
is indeed an instable saddle point, hence the sphaleron of the basic model 
(\ref{m1}).

It is a special feature of the geometrical NCL construction that the
spherically symmetric loop configuration (\ref{m14}) involves only a Higgs 
triplet ($i=1,2,3$) and an $so(3)$ gauge field. This ``symmetry breakdown'' 
along the loop, together with the fact the the static $SO(4)$ Higgs 
Hamiltonian (\ref{m6}) is formally equal to the generalised $SO(3)$ Higgs
Hamiltonian of Ref.\cite{monopole} eq.\ (3) under the coupling constant mapping
\begin{equation}
\lambda_a\mapsto \frac{1}{a}\lambda_a, \quad a=1,2,3,4,
\end{equation}
justifies the interpretation that the generalised $SO(3)$
Higgs monopole \cite{monopole} is really ``embedded'' into the $SO(4)$
Higgs model under consideration, gaining its instability from the additional
gauge degrees of freedom which are excited along the loop, an effect which 
also occurs in lower dimensional $SO(d)$ Higgs models with instanton and
sphaleron \cite{d23}.

\section{NCL, sphaleron and Chern--Simons number}

The construction of the geometrical loop in the previous section
was guided by the topological properties of static finite energy
configurations. The sphaleron itself  however, is not 
a ``topological object''. It can be visualised as top of
the static energy barrier separating topologically distinct vacua,
whereas the instanton interpolates between these vacua in Euclidean 
spacetime. This instanton is topologically stable due to the existence 
of a lower bound to the Euclidean action (\ref{m1}),
\begin{equation}
\int \CL\,d^4 x
\ge\min\{1,\lambda_a\}\lim_{R\rightarrow\infty}\int_{S^3(R)}\Omega
\label{l1}
\end{equation}
($R^2=|x_{\mu}|^2$)
where the Chern--Simons form $\Omega$, which results from the dimensional
reduction of the {\it fourth} Chern class, is given by 
$\Omega=\Omega^{(2)}+\Omega^{(1)}+\Omega^{(0)}$,
\begin{eqnarray}
\Omega^{(2)} & = & -\frac12 \eta^4 
\tr\left(\gamma_5A_{\nu}\left(F_{\rho\sigma}-
\frac23A_{\rho}A_{\sigma}\right)
\right)\d x^{\nu}\wedge \d x^{\rho}\wedge \d x^{\sigma} \nonumber \\
\Omega^{(1)} & = & \frac{1}{12} \eta^2 
\tr\left(\gamma_5\Phi S_{\nu\rho\sigma}\right)
\d x^{\nu}\wedge \d x^{\rho}\wedge \d x^{\sigma} \nonumber \\
\Omega^{(0)} & = & \frac{i}{36} 
\tr\left(\gamma_5\Phi S_{[\nu\rho}D_{\sigma]}\Phi\right)
\d x^{\nu}\wedge \d x^{\rho}\wedge \d x^{\sigma}.
\label{l2}
\end{eqnarray}
This topological bound allows the classification of the instantons
of the model in terms of integer Chern--Pontryagin charge, 
\begin{equation}
q=-\frac{1}{8\pi^2\eta^4}\lim_{r\rightarrow\infty}\int_{S^3(r)}\Omega
=-\frac{1}{8\pi^2\eta^4}\int_{\R^4}\d\Omega.
\label{l3}
\end{equation}

\begin{figure}
\begin{center}
\includegraphics[bb=3cm 10cm 18cm 21cm,angle=0,scale=0.8]{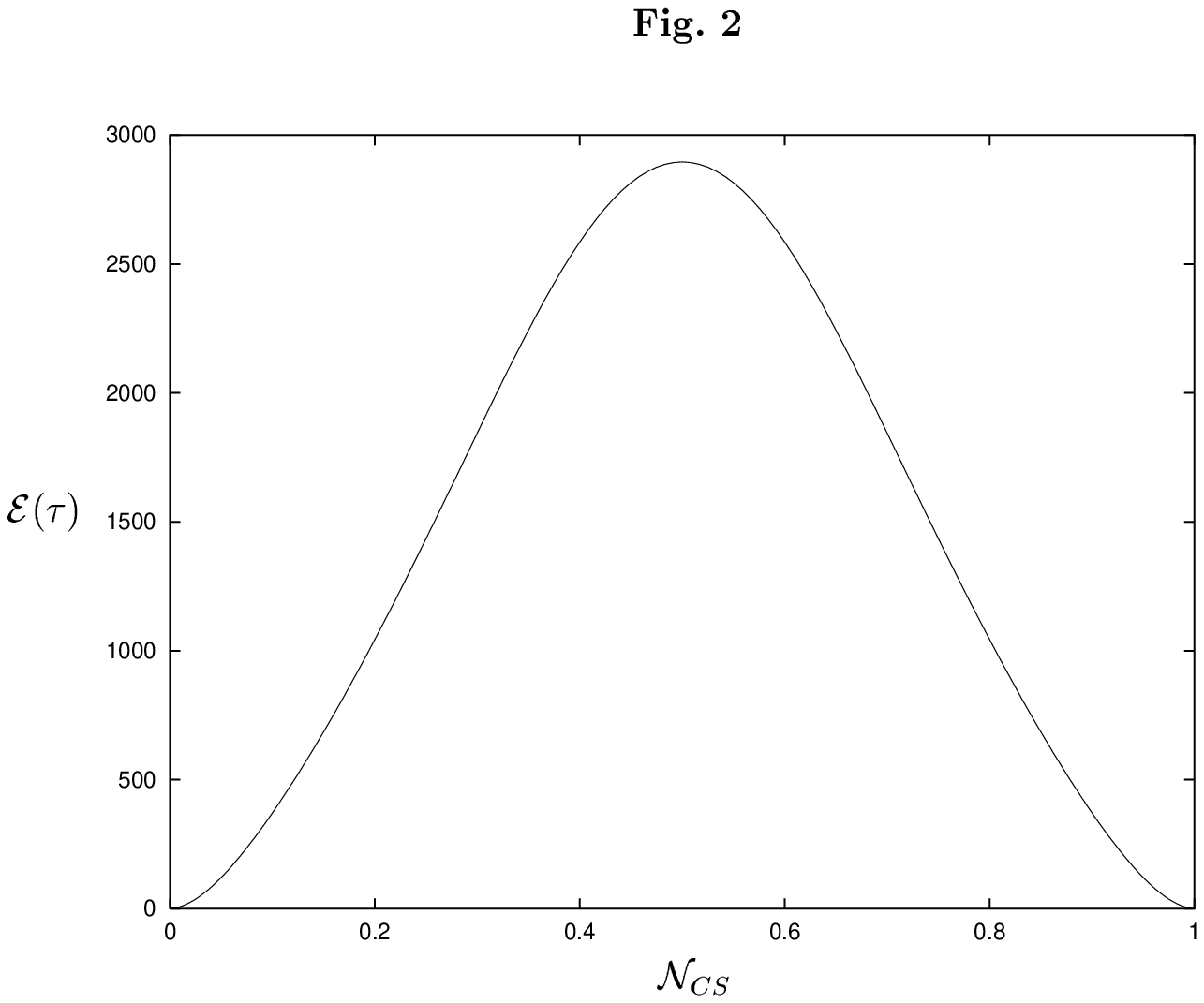}
\end{center}
\caption{Energy along the geometrical NCL in terms of increasing Chern--Simons 
number $\Ncs$, $\lambda_a=\frac{1}{a}$}
\label{so4fig2}
\end{figure}

The connection between instantons and sphalerons as objects relating 
topologically distinct vacua becomes obvious if one interprets the 
NCL connecting the vacua through the sphaleron as
an object in Euclidean spacetime, treating the loop parameter as time $t=x_0$
depend $\tau=\tau(t)$ such that $\tau(t\rightarrow\infty)=0$,
$\tau(t\rightarrow-\infty)=\pi$. Splitting the spacetime integral
(\ref{l3}) into two parts, consisting of a spatial surface-- and a 
spatial volume--integral  yields what is called 
the Chern--Simons number at time $t_0$,
\begin{equation}
\Ncs(t_0)=-\frac{1}{8\pi^2\eta^4}\left[\int_{\R^3}
\Omega_0\bigg|_{t=-\infty}^{t=t_0}+
\int_{-\infty}^{t_0}dt\lim_{r\rightarrow\infty}\int_{S^2(r)}\hat{\Omega}\right]
\label{l4}
\end{equation}
($r^2=|x_i|^2$)
where the three--form $\Omega_0=\Omega_0^{(2)}+\Omega_0^{(1)}+\Omega_0^{(0)}$
and the two--form $\hat{\Omega}=\hat{\Omega}^{(2)}+\hat{\Omega}^{(1)}+
\hat{\Omega}^{(0)}$ in spherical coordinates are found to be
\begin{eqnarray}
\Omega_0^{(2)} & = &
\eta^4\tr\left(\gamma_5\left(F_{[r\theta}A_{\phi]}+
\frac23A_{[r}A_{\theta}A_{\phi]}\right)\right)
\d r \wedge \d\theta \wedge \d\phi
\nonumber \\
\Omega_0^{(1)} & = & 
\frac{1}{2}\eta^2\tr\left(\gamma_5\Phi S_{r\theta\phi}\right)
\d r \wedge \d\theta \wedge \d\phi
\nonumber \\
\Omega_0^{(0)} & = & 
\frac{i}{6} \tr\left(\gamma_5\Phi S_{[r\theta}D_{\phi]}\Phi\right)
\d r \wedge \d\theta \wedge \d\phi \nonumber \\
\hat{\Omega}^{(2)}
& = &  - \eta^4\tr\left(\gamma_5\left(F_{[0\theta}A_{\phi]}+
\frac23A_{[0}A_{\theta}A_{\phi]}\right)\right) \d\theta \wedge \d\phi
\nonumber \\
\hat{\Omega}^{(1)}
& = &  -\frac{1}{2}\eta^2\tr\left(\gamma_5\Phi S_{0\theta\phi}\right)
\d\theta \wedge \d\phi
\nonumber \\
\hat{\Omega}^{(0)} 
& = & -\frac{i}{6} \tr\left(\gamma_5\Phi S_{[0\theta}D_{\phi]}\Phi\right)
\d\theta \wedge \d\phi.
\label{l5}
\end{eqnarray}
Integrating over infinite time, the Chern--Simons number equals the 
Chern--Pontryagin charge, $\Ncs(t_0=\infty)=q$.

Assuming $\Omega_0(t\rightarrow-\infty,\vec{x})=0$  which fixes 
$\Ncs(t\rightarrow -\infty)=0$ for the initial vacuum,
the volume and surface contributions to the Chern--Simons number along the 
loop, respectively, are then found to be
\begin{eqnarray}
& &\int_{\R^3} \Omega_0  =
\frac{8\pi\eta^4}{3}\sin^3\tau\cos\tau\int_0^{\infty} 
\left\{(h^2-3-\sin^2\tau)\left[(1-f^2h'-2(1-h)f'W\right]
+6h'W^2\right\}dr, \nonumber \\
& &\int_{-\infty}^{t_0}dt\lim_{r\rightarrow\infty}\int_{S^2(r)}\hat{\Omega} =
-8\pi\eta^4\left[\tau(t_0)-\frac12\sin 2\tau(t_0)\right].
\label{l6}
\end{eqnarray}
In particular, the volume integral contribution vanishes if the loop reaches
the sphaleron at time $t_0=t_S$, $\tau(t_S)=\frac{\pi}{2}$, and the 
``Chern--Simons number of the sphaleron'' is found to be $\Ncs(t_S)=\frac12$. 
At infinite time, $t_0\rightarrow\infty$, the volume contribution vanishes, and
one immediately finds $\Ncs(t_0\rightarrow\infty)=1$, relating the NCL 
interpreted as Euclidean spacetime configuration to the instanton.

The volume integral can be computed numerically, using the monopole profile
functions $(h^{(M)},f^{(M)})$ which were also used to evaluate the energy
along the loop in the previous section. This allows to plot the energy
barrier between the two topologically distinct vacua in terms of increasing
Chern--Simons number as shown in Fig.\ \ref{so4fig2}.

\section{Summary and discussion}

We have presented the complete construction of the sphaleron solution to
the model (\ref{m1}), together with a finite energy barrier NCL. This
completes the verification that this model supports both finite size
instantons and sphalerons. Apart from the relevance of this to the
study of the periodic instantons~\cite{KRT,KT}, some extensions of this
model are of some physical relevance, which we now discuss briefly.
The results obtained here remain qualitatively unchanged under the
extensions discussed below.

The Lagrangian (\ref{m1}) can be called the {\em ``basic model''}, and 
is derived from the 
eight--dimensional generalised Yang--Mills system by dimensional reduction
\cite{T}. For physical reasons, it is convenient to consider an 
{\em ``extended model''}, adding the term
\begin{equation}
\CL_{ext}=\tr\left(-\mu_1(D_{\mu}\Phi)^2+\mu_2S^2\right)
\label{m5}
\end{equation}
to the Euclidean Lagrangian (\ref{m1}). The main effect of adding
(\ref{m5}) to the {\em ``basic model''} is to force a time-independent
vacuum field, e.g.\
$\Phi_{\rm {vac}}=\eta \gamma_5 \gamma_4$. In Ref.~\cite{AOT}
another extended version incorporating the term
$\tr F_{\mu \nu}F_{\nu \lambda}F_{\lambda \mu}$ was considered, to enable
the construction of a Coulomb gas of instantons. It is interesting to
note that in the absence of (\ref{m5}), the instanton now
would be power localised, the exponential localisation being restored
in the presence of (\ref{m5}).

Another interesting effect of adding (\ref{m5}) to (\ref{m1}) is the
resulting spectrum when the Higgs mechanism is applied. Using the
notation $\vec \gamma =(\gamma_{\alpha},\gamma_4)$, $\alpha =1,2,3$, and
expanding around the Higgs vacuum $\eta \gamma_5 \gamma_4$
\begin{equation}
\label{vac}
\Phi =\gamma_5 [\gamma_4 (\eta + v) +\zeta_{\alpha} \gamma_{\alpha}] \: ,
\end{equation}
there appears a mass-like term for an isovector vector field in the
(Minkowskian) Lagrangian. It is the $so(3)$ part of the gauge field 
corresponding to
the broken $SO(3)$ subgroup of the gauge group. This part of the $so(4)$
gauge field fluctuation, $W_{\mu}^{\alpha}=2A_{\mu}^{\alpha 4}$ in
\[
A_{\mu}=A_{\mu}^{\alpha \beta}\gamma_{\alpha \beta}+
W_{\mu}^{\alpha}\gamma_{\alpha 4} \: ,
\]
consists of the components that do not commute with $\Phi_{\rm{vac}}$. The
Higgs field then has one component that stays massive, described by the
scalar field $v$ in (\ref{vac}) and the corresponding Goldstone Bosons
are swallowed via the gauge $SO(3)$ transformation
\begin{equation}
\label{gauge}
g= {\rm exp}\frac{1}{\eta} \gamma_{\alpha 4} \zeta_{\alpha} \: ,
\end{equation}
leaving the mass--like term $\mu_1 \eta^2 W_{\mu}W^{\mu}$ in the Lagrangian.
Strictly speaking this term is not a mass term in the sense that the
accompanying quadratic kinetic term
$(\partial_{\mu}W_{\nu}-\partial_{\nu}W_{\mu})
(\partial^{\mu}W^{\nu}-\partial^{\nu}W_{\mu})$ is absent. This is the
result of the absence of the usual Yang-Mills term
$\mbox{tr}F_{\mu \nu}^2$ in the Lagrangian (\ref{m1}).

This brings us to the final item of discussion, namely the question of the
absence of the YM term $\mbox{tr}F_{\mu \nu}^2$ in (\ref{m1}).
Independently of the dynamics of the $SO(4)$ system with the Higgs field
in the 4-vector representation, there exists a Dirac gauge in which the
Higgs field can be gauged to a constant at infinity. It follows that
the asymptotic $so(4)$ gauge field must decay with the inverse power of
$r$, with an inverse--square decay of the curvature. In 4 dimensions, the
contribution of the YM term $\mbox{tr}F_{\mu \nu}^2$ in (\ref{m1}) 
to the action will
then be logarithmically divergent. Thus, the most important property of
this model (and its extended versions), namely its suitability for
describing a Coulomb gas of instantons, prevents the presence of the
usual YM term with the consequence that the gauge field and its massive
component are not endowed with a propagator. 

To complete this argument one has to eliminate the following possibility:
Namely that instead of exploiting the descendent of the {\it fourth}
Chern--Pontryagin charge (\ref{l2}), one exploits the {\it second}
Chern--Pontryagin charge $\mbox{tr}F_{\mu \nu}\: ^{\star}F_{\mu \nu}$.
In that case, the topological lower bound
\begin{equation}
\label{bound}
\mbox{tr}\: (F_{\mu \nu}^2 +{\rm Higgs\: terms})\: \: \ge
\mbox{tr}\: (F_{\mu \nu}\: ^{\star} F_{\mu \nu})
\end{equation}
holds and as a result the curvature strength decays as the inverse power
of $r^4$, corresponding to an asymptotically pure gauge connection like
the BPST~\cite{BPST} instanton. The problem here is that the finite action
condition for the Higgs terms in (\ref{bound}) results in an asymptotic
connection field that decays as {\it one half} times a pure gauge, which is
inconsistent with the finite action requirement on the YM terms, namely
that the connection be asymptotically pure gauge.


\begin{thebibliography}{10}

\bibitem{OT1}
G.M. O'Brien and D.H. Tchrakian, Mod.\ Phys.\ Lett.\ {\bf A4} (1989) 1389.

\bibitem{AOT}
K. Arthur, G.M. O'Brien and D.H. Tchrakian, J. Math.\ Phys.\ {\bf 38} (1997)
4403.

\bibitem{T}
see for example, D.H. Tchrakian, Yang-Mills Hierarchy, in Proceedings of
XXI International Conference on Differential Geometric Methods in
Theoretical Physics, Int.\ J. Mod.\ Phys.\ A (Proc. Suppl.) {\bf 3A} (1993)
584.

\bibitem{OT2}
G.M. O'Brien and D.H. Tchrakian, Phys.\ Lett.\ {\bf B 282}  (1992) 111.

\bibitem{KOT}
B. Kleihaus, D. O'Keeffe and D.H. Tchrakian, Phys.\ Lett.\
{\bf B 427} (1998) 327.

\bibitem{Manton}
N.S. Manton, Phys.\ Rev.\ D {\bf 28} (1983) 2019; 
F.R. Klinkhamer and N.S. Manton, 
Phys.\ Rev.\ D {\bf 30} (1984) 2212.

\bibitem{P}
A.M. Polyakov, Nucl.\ Phys.\ B {\bf 120} (1977) 429.

\bibitem{A}
I. Affleck, Nucl.\ Phys.\ B {\bf 191} (1981) 429; I. Affleck, M. Dine and N.
Seiberg, {\it ibid.} B {\bf 241} (1984) 493; {\it ibid.} B {\bf 256}
(1985) 557.

\bibitem{KRT}
S.Yu. Khlebnikov, V.A. Rubakov and P.G. Tinyakov, Nucl.\ Phys.\ {\bf 367}
(1991) 334.

\bibitem{KT}
A.N. Kuznetsov and P.G. Tinyakov, Phys.\ Lett.\ B {\bf 406} (1997) 76.

\bibitem{BPST}
A.A. Belavin, A.M. Polyakov, A.S. Schwarz and Yu.S. Tyupkin, Phys.\ Lett.\
B {\bf 59} (1985) 85.

\bibitem{tHP}
G.'tHooft, Nucl. Phys. {\bf B 79} (1974) 276; A.M. Polyakov, JETP Lett.\
{\bf 20} (1974) 194.

\bibitem{monopole} B. Kleihaus, D. O'Keeffe and D.H. Tchrakian,
Phys.\ Lett.\ {\bf B427}(1998)327. 

\bibitem{d23} B. Kleihaus, D.H. Tchrakian and F. Zimmerschied, 
``$d$--dimensional $SO(d)$--Higgs models with instantons and sphaleron: 
$d=2,3$'', hep--th/9904048

\end{thebibliography}
\end{document}